\documentclass{article}

\usepackage{spconf,amsmath,graphicx}
\usepackage{multirow}
\usepackage{booktabs}       % professional-quality tables
\usepackage{url}
\usepackage{algorithm}
\usepackage{algpseudocode,color}
\usepackage{float} % here for H placement parameter
\usepackage{arydshln}
\usepackage{etoolbox}
\usepackage{graphicx}
\usepackage{subfigure}

\algnewcommand\algorithmicforeach{\textbf{for each}}
\algdef{S}[FOR]{ForEach}[1]{\algorithmicforeach\ #1\ \algorithmicdo}

\newtoggle{comments}
\newtoggle{blind}
\newtoggle{highlightupdates}

\toggletrue{comments}
\toggletrue{blind}
\toggletrue{highlightupdates}

\title{Hierarchical Recurrent Adapters for Efficient Multi-Task Adaptation of Large Speech Models}

\name{\parbox{.95\linewidth}{\centering
Tsendsuren Munkhdalai, Youzheng Chen, Khe Chai Sim \\ Fadi Biadsy, Tara Sainath, Pedro Moreno Mengibar}}
\address{Google, USA}

\begin{document}
% \ninept
\maketitle
\begin{abstract}

Parameter efficient adaptation methods have become a key mechanism to train large pre-trained models for downstream tasks. However, their per-task parameter overhead is considered still high when the number of downstream tasks to adapt for is large. We introduce an adapter module that has a better efficiency in large scale multi-task adaptation scenario. Our adapter is hierarchical in terms of how the adapter parameters are allocated. The adapter consists of a single shared controller network and multiple task-level adapter heads to reduce the per-task parameter overhead without performance regression on downstream tasks. The adapter is also recurrent so the entire adapter parameters are reused across different layers of the pre-trained model. Our Hierarchical Recurrent Adapter (HRA) outperforms the previous adapter-based approaches as well as full model fine-tuning baseline in both single and multi-task adaptation settings when evaluated on automatic speech recognition tasks.

% (Requirement: within 1000 chars):
\end{abstract}

\noindent\textbf{Index Terms}: large pre-trained models, parameter efficient adaptation, recurrent neural networks
\section{Introduction}
\label{sec:introduction}

There has been a paradigm shift towards adapting a single large pre-trained model to multiple downstream tasks. Full model adaptation such as fine-tuning is expensive as the entire model specializes on a single task~\cite{bert}. Since the per-task parameter overhead becomes as large as all model weights, the full fine-tuning approach is not scalable in applications with a large number of tasks, like personalized speech recognition~\cite{sim2021robust,pundak2022fly,munkhdalai2023nam+}.

Parameter efficient adaptation methods on the other hand focus on fine-tuning a fraction of model weights (i.e. the final dense layer before softmax) or adding a small number of task specialized parameters. There are two main categories of parameter efficient adaptation of large pre-trained models: soft-prompt tuning and the adapter methods. Adapter layers have shown better performance on a variety of tasks, thanks to its high computational capacity and more parameters. On the other hand, the soft-prompt tuning approaches offer a more flexible, efficient way to adapt and deploy large models as it is straightforward to use the soft prompt vectors for mixed-task batches during inference. However, the capability of the current prompt tuning techniques are limited by the capacity of the prompt vectors. Optimizing them via back-propagation is not a straightforward procedure. As a result, they under-perform on harder text generation tasks, like machine translation and summarization~\cite{an2022input}. 
Furthermore, it is unclear how to combine the existing soft-prompt tuning techniques with streaming speech models due to the changing input and attention window.

\begin{figure}[t]
    \begin{minipage}[b]{1.0\linewidth}
      \centering
      \centerline{\includegraphics[scale=0.6]{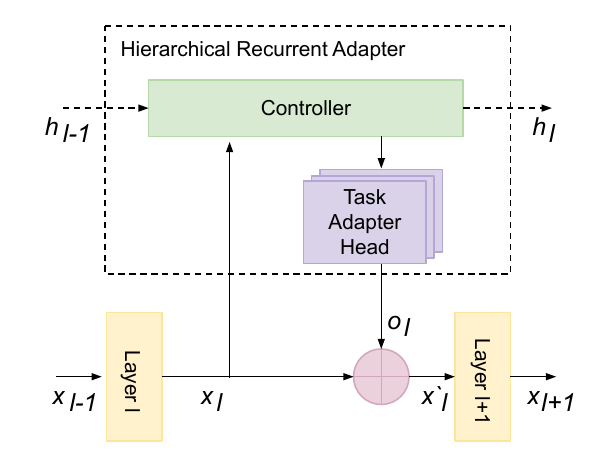}}
    \end{minipage}
    \caption{Hierarchical Recurrent Adapter (HRA). The yellow box indicates layers of the underlying backbone speech model. The HRA consists of a single recurrent controller and multiple task-level adapter heads. The output of the adapter head is added to the backbone feature for adaptation of downstream speech tasks. In HRA, the adapter heads and the recurrent controller weights are shared across all layers keeping the adapter parameter overhead minimal.}
    \label{fig:hra}
\end{figure}

In this work, we focus on parameter efficient adapter methods for adaptation of large pre-trained speech models for automatic speech recognition (ASR) tasks. There is a line of works on efficient adapters, including residual adapters~\cite{rebuffi2017learning}, Low-Rank Adapter (LoRA)~\cite{hu2021lora}, and BitFit~\cite{zaken2021bitfit}. The residual adapters incorporate a 2-layer feed-forward network (FFN) as adapter for each pre-trained Transformer~\cite{vaswani2017attention} or Conformer~\cite{gulati2020conformer} block. The adapter can be placed in parallel or sequential to an entire block or the FFN layers within the block. It utilizes a hidden layer bottleneck to reduce the number of parameters and avoid over-fitting on a small downstream task data. Despite the simplicity, Residual Adapters have been successfully applied to many NLP, speech and vision tasks~\cite{rebuffi2017learning,tomanek2021residual,li2023modular}. 

LoRA~\cite{hu2021lora} is a more recent adapter approach that decomposes the adapter matrix into two low-rank matrices for better parameter efficiency and learn a task-parameter difference for the downstream tasks, similar to MetaNet with Fast-Weight adapters~\cite{munkhdalai2017meta,munkhdalai2020sparse}. In LoRA, the task specific weight matrix can be recovered by multiplying the two small decomposing matrices and the adapter matrices can be added next to any weight matrix. BitFit~\cite{zaken2021bitfit} on the other hand adds no additional parameters and fine-tunes only the bias and scaling vector terms for a new task. Another concurrently developed work is READ that applies a recurrent neural network as adapter for parameter and computation efficiency~\cite{wang2023read}. READ was introduced for adaptation of Large Language Models and focuses on NLP tasks. This approach is also related to feature fusion methods that aims to provide more efficient training~\cite{huo2023resource}.

The existing adapter methods were mainly developed for single or few task adaptations settings; and thus their per-task parameter overhead is high in large scale multi-task scenario. To reduce the per-task parameter overhead, we introduce a hierarchical adapter approach dubbed Hierarchical Recurrent Adapter (HRA). HRA is equipped with a recurrent controller network and a set of task-level adapter heads. The recurrent controller network is shared across all tasks while the task-level adapter head is specialized for each task. Since HRA is recurrent along the depth of the large pre-trained model, HRA parameters are shared across the layers as well. Therefore, the per-task parameter overhead becomes only task-level adapter head. 

In our extensive experiment on ASR, we show that our HRA achieves better WERs with 2-8x less parameters in single-task as well as multi-task evaluations. The HRA closes the WER gap against the full fine-tuning baseline and improves further.

The contribution of this work is 3-fold. First, we show that an improved model-wise parameter efficiency is achieved by adapter recurrency. Second, this work introduces a modular adaptation model by decomposing the adapter module into controller network and task adapter heads. Finally, we achieve a better task-wise parameter efficiency via the adapter heads.

\section{Methods}

As shown in Figure~\ref{fig:hra}, the proposed Hierarchical Recurrent Adapter consists of a single shared controller and multiple task specific adapter heads. We add one adapter head per task. Only the adapter head parameters are trained when there is new task coming in. We experiment with two simple adapter head architecture: simple linear projection and FFN heads. 

The shared controller is responsible for interacting with task specialized adapter heads.
Furthermore, unlike residual adapters our HRA is shared across all layers of a pre-trained large model to keep adapter parameters small.
We provide a detailed description of each component below.

\subsection{Recurrent Controller}
The controller is shared for all layers of the underlying backbone model as well as tasks and is responsible for orchestrating the interaction between the backbone model and task specialized adapter heads. The controller takes in as input the activation $x_{l}$ at layer $l$ of the backbone model and computes a new interaction recurrent vector $h_l$ for task-level adapter. Since the controller is a recurrent network, it also takes in its last hidden activation $h_{l-1}$.

We choose to parameterize our adapter controller with a lightweight recurrent network for parameter and inference efficiency. Specifically, we use IndRNN~\cite{li2019deep} as it is computationally cheaper than the other RNN variants and admits ReLU function as its activation without a gradient explosion issue. IndRNN computes its recurrent activation $h_l$ as:

\begin{equation}
\label{eq:indrnn}
h_l = ReLU(W x_l + u h_{l-1} + b)
\end{equation}
where $x_l$ is the RNN input feature representation extracted from the $l^{th}$ layer of the backbone speech model and $W$, $u$ and $b$ are input projection matrix, recurrent scaling vector and the bias term.

\subsection{Task Adapter Heads}
Once the new interaction recurrent vector $h_l$ is computed as in Eq~\eqref{eq:indrnn}, we learn an adapter output $o_l$ for backbone layer $l$ by passing it through the task-level adapter head. The adapter output $o_l$ is then added back to the original feature activation to obtain task-specific representation $x'_l$:

\begin{equation}
x'_l = x_l + o_l.
\end{equation}

The resulting representation $x'_l$ is further given as input to the next backbone layer $l+1$.

Similar to the controller, the task adapter head is also shared across the layers of the backbone model resulting in a compact HRA adapter for all tasks. We consider linear project matrix and a 2-layer FFN for the adapter head.

\subsubsection{Linear Adapter Head}
We can use a simple linear projection matrix as task-level memory, so to adapt to a new task we incorporate and fine-tune only a single linear projection matrix. Given the controller hidden state $h_l$ the linear projection head then computes the output $o_l$ as:

\begin{equation}
o_l = M_n h_l
\end{equation}
where $M_n$ is the task-specific project matrix and $n$ is the task index.

\subsubsection{Feed-Forward Adapter Head}
We can apply a 2-layer FF neural network with ReLU activation as the task-level adapter head. In this case, the adapter output is computed as:

\begin{equation}
o_t = M_{2,n} ReLU(M_{1,n} h_t)
\end{equation}
where $M_{2,n}$ and $M_{1,n}$ are the task-level head weights for the $n^{th}$ task.

\section{Experimental Setup}
\label{sec:experiments}
We run two sets of experiments. One focuses on the evaluation of single-task adaptation performance of our proposed HRA adapters and the other is on the multi-task adaptation scenario. For the single-task evaluation, we used a multi-domain corpora as training and voice-search (VS) dataset as test. We also evaluated each model on a harder VS test set with proper nouns like person names.

For the multi-task setup, we use 
Euphonia corpora
% a dysarthric speech corpora [name anonymized for blind review purposes]
, atypical speech dataset consisting of over 1 million utterance recordings of over 1000 anonymized speakers with different types and severity levels of speech impairments. 

\subsection{Pre-trained Model}

We started with a pre-trained Universal Speech Model (USM) ~\cite{zhang2023google}. This model has ~2 billion parameters and was pre-trained with the BEST-RQ objective~\cite{chiu2022self} on large unlabeled multilingual corpora of 12 million hours covering over 300 languages.
We then apply different adapter techniques to the pre-trained USM model for adaptation of ASR tasks.
The adapter methods as well as full model fine-tuning baseline are trained by using the CTC loss~\cite{graves2006connectionist} for ASR.

\subsection{Datasets}
\label{sec:dataset}

All collected experimental data sets adhere to the Privacy Principles in~\cite{privacyprinciples} and AI Principles in~\cite{aiprinciples}.

\subsubsection{Multi-domain Corpora}
The multi-domain corpora was used to train the adapter parameters in single-task evaluation experiments. It~\cite{narayanan2018toward} consists of anonymized English utterances from domains including voice search, far-field and long-form. The speech transcripts contain a mix of human-transcribed labels and machine-transcribed labels produced by teacher ASR models~\cite{hwang22c_interspeech}.

\subsubsection{Euphonia corpora}
% \subsubsection{Dysarthric Speech Corpora [name anonymized for blind review purposes]}
We carefully select 128 speakers with speech impairments from the dysarthric speech [name anonymized for blind review purposes] corpus
~\cite{macdonald2021disordered}
, including speakers with ALS, Down-Syndrome, Cerebral Palsy, Parkinson's Stroke, and other etiologies. Recording text prompts consists of a variety of domains, such as caregiver phrases, conversational sentences, movie quotes, and assistant phrases. We split 80\% for train, 10\% for cross-validation and 10\% for test on each speaker based on transcript, and there is no transcript overlapping among train set, cross-validation set, and test set. Speaker identifiers are provided along with each data utterance. We separate the test set into 128 sub sets so that each one only consists of one speaker for evaluation purposes.

\section{Results}

\begin{table}[t]
    \centering
    \caption{Single-task adaptation WER results on voice-search (VS) and voice-search with proper nouns (VS w. PN) test sets. \# Params. row shows the number of adapter parameters. Our FFN Head HRA outperforms the full fine-tuning baseline at 12.8M parameters.
    }
    \label{tab:single-task}
    \begin{tabular}{l c c c}
        \toprule
        \multirow{1}*{Model} & \# Params. & VS & VS w. PN \\
        \toprule
        \multirow{1}*{Full Fine-tuning}
        &    1.8B    &  5.3    &    15.7  \\
        \midrule
        \multirow{1}*{BitFit}
        &    1.3M    &  6.6    &    18.4  \\
        \midrule
        \multirow{3}*{LoRA}
        &   2.0M & 7.5 &	19.9 \\
        &   4.0M & 6.8 &	19.0 \\
        &   7.9M & 6.4 &	18.0 \\
        \midrule
        \multirow{3}*{Residual Adapters}
        &   3.2M & 6.3 &	17.9 \\
        &   6.4M & 6.2 &	17.1 \\
        &   12.7M & 5.8 &	16.7 \\
        \midrule
        % \multicolumn{4}{c}{Ours}
        % \bf Ours & & & \\
        \midrule
        \multirow{3}*{Linear Head HRA (ours)}
        &   814K & 6.2 &	17.4 \\
        % &   1.6M & 6.0 &	16.9 \\
        % &   3.2M & 5.7 &	16.7 \\
        &   6.4M & 5.4 &	16.2 \\
        &   12.8M & \bf 5.1 &	15.7 \\
        \midrule
        \multirow{3}*{FFN Head HRA (ours)}
        &   1.3M & 6.0 &	17.1 \\
        % &   2.9M & 5.9 &	16.6 \\
        % &   6.8M & 5.5 &	16.1 \\
        &   13.6M & 5.2 &	15.4 \\
        &   27.2M & \bf 5.1 &	\bf 15.3 \\
        \bottomrule
    \end{tabular}
\end{table}

\subsection{Single-task Adaptation}
\label{sec:single-task}
Table~\ref{tab:single-task} reports the WER results from our single-task adaptation experiments. Unless otherwise mentioned, all models were trained for 100K iterations. 

As for the baselines, we trained a full model fine-tuning as well as other adapter techniques, such as BitFit, LoRA and Residual Adapters. For LoRA, we set its low-rank hyper-parameter to be 4, 8 and 16. We varied the Residual Adapter bottleneck dimension to be 32, 64 and 256 and recurrent dimension of HRA to 256, 2048 and 4096.

The despite the simplicity, BitFit obtains a strong WER of 6.6 on VS test set while LoRA seems to beat BitFit with a WER of 6.4 only after ~8M parameters. The Residual Adapters on the other hand show robust results across different adapter sizes and the more adapter parameters improve the WER.

The last two sets of rows present our HRA results. Our smallest adapter - the HRA with Linear Head can achieve 6.2 WER at 814K parameters and this WER is already better than BitFit, all LoRA and smaller Residual Adapter results. This adapter matches the WER of the Residual Adapter with 6.4M parameters, showing 8x parameter efficiency. Our Linear Head HRA with 12.8M parameters already outperforms the full fine-tuning baseline and the largest FFN Head HRA further sets a new state-of-the-art WER on both test sets.

\begin{figure}[t]
    \begin{minipage}[b]{1.0\linewidth}
      \centering
      \centerline{\includegraphics[scale=0.45]{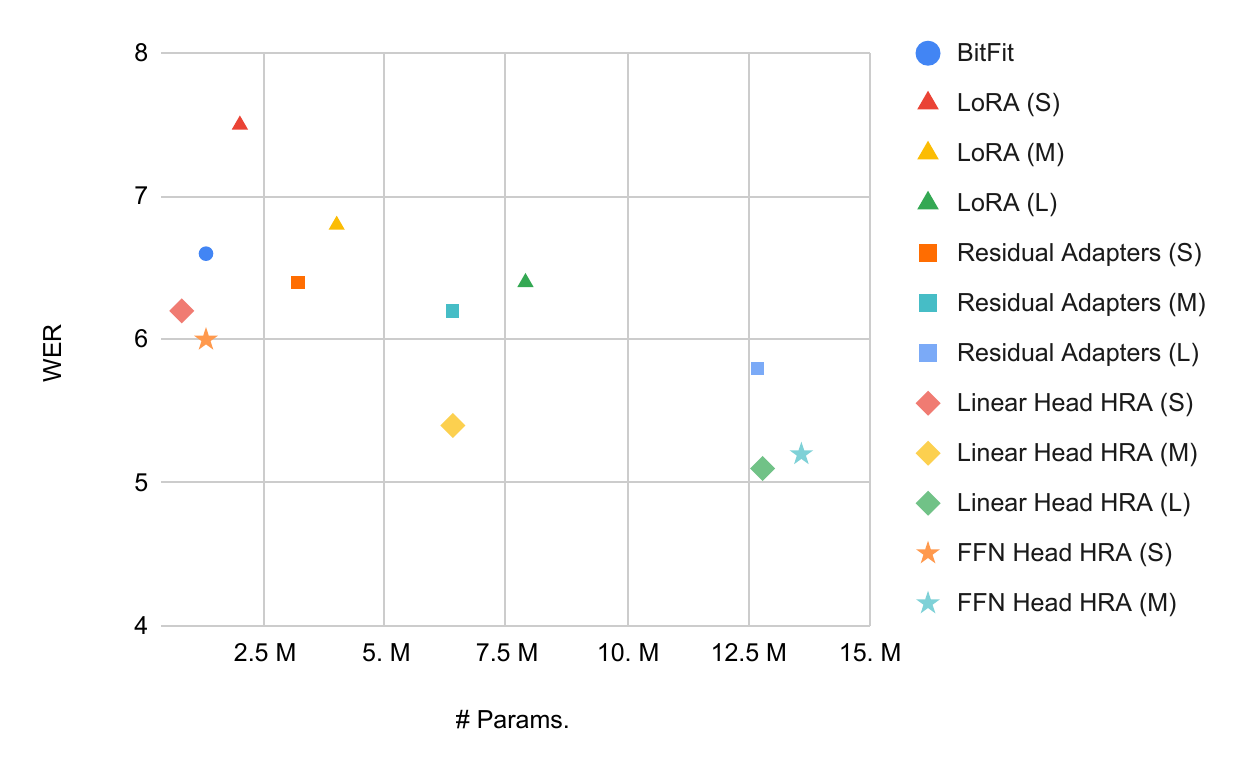}}
    \end{minipage}
    \caption{Parameter efficiency of different adapter methods. Lower-left points are more parameter efficient (x-axis is truncated at 15M).}
    \label{fig:parameter_eff}
\end{figure}

% Our FFN Head HRA (L) with 6.8M parameters achieves the best WER on both VS and VS w. PN test sets while having ~2x smaller parameter than the second best model - Residual Adapters (L) with 12.7M parameters. This adapter further closes the gap between the full fine-tuned model and adapters and the relative WER difference is now less than ~4\% on VS test set.

In Figure~\ref{fig:parameter_eff}, we plotted the WER against the number of adapter parameters. The lower-left points represent more parameter efficient methods as both WER and the number of parameters are lower simultaneously and we can see that HRA methods are mainly clustered around that region.

\begin{table}[t]
    \centering
    \caption{Multi-task adaptation WER results on
    Euphonia
    % dysarthric speech
    data sets. Our FFN Head HRA achieves the best WER and closes the gap against full fine-tuning  baseline. Figure~\ref{fig:param_task} shows that this model has a sub-linear growth in terms of the size of adapter parameters as the number of tasks increases.}
    \label{tab:multi-task}
    \begin{tabular}{l c c c c}
        \toprule
        \multirow{1}*{Model} & \# Params. & Mean & Median & SD \\
        \toprule
        \multirow{1}*{USM Basemodel}
        & - &    31.5    &  21.8    &    28.6  \\
        \midrule
        \multirow{1}*{Full Fine-tuning}
        & 232B &    9.3    &  5.4    &    11.1  \\
        \midrule
        \multirow{3}*{LoRA}
        & 201M &   10.9 & 6.6 &	12.4 \\
        & 403M & 10.9 & 7.4 & 11.6 \\
        & 805M & 12.4 & 6.9 & 15.8 \\
        \midrule
        \multirow{3}*{Residual Adapters}
        & 410M &   10.2 & 6.1 &	11.6 \\
        & 819M & 10.2 & 6.1 & 11.2 \\
        & 1.6B & 10.1 & 6.2 & 11.0 \\
        \midrule
        \midrule
        \multirow{3}*{Linear Head HRA}
        & 51M & 14.6 & 9.7 & 14.2 \\
        & 102M & 14.5 & 9.9 & 13.1 \\
        & 203M & 16.1 & 12.0 & 12.1 \\
        \midrule
        \multirow{3}*{FFN Head HRA}
        & 201M & \bf 9.9 & 6.3 &	11.2 \\
        & 403M & 10.2 & 6.1 & 11.8 \\
        & 806M & 10.4 & 6.2 & 11.3 \\
        \bottomrule
    \end{tabular}
\end{table}

\subsection{Multi-task Adaptation}
\label{sec:multi-task}

Table~\ref{tab:multi-task} reports the WER results from our multi task adaptation experiments. We build golden baseline from USM model with full model fine-tuning on each speaker respectively, and each model is fine-tuned with data from its corresponding speaker only. For the adapter configurations, we parameterize adapters by a speaker-id and learnable one-hot embedding. Following \cite{biadsy2022scalable}, we introduce one-hot-embedding lookup table with entries through one-on-one mapping to corresponding speakers. During training, we randomly select data samples from the 128 speakers in each batch. The recurrent controller network is shared across all 128 speakers while a separate adapter head is inserted for each speaker for specialization. For adapter baseline, we choose to experiment with LoRA and Residual Adapters since it showed a promising performance in the single-task adaptation setup (Section~\ref{sec:single-task}).

\begin{figure}[H]
    \begin{minipage}[b]{1.0\linewidth}
      \centering
      \centerline{\includegraphics[scale=0.45]{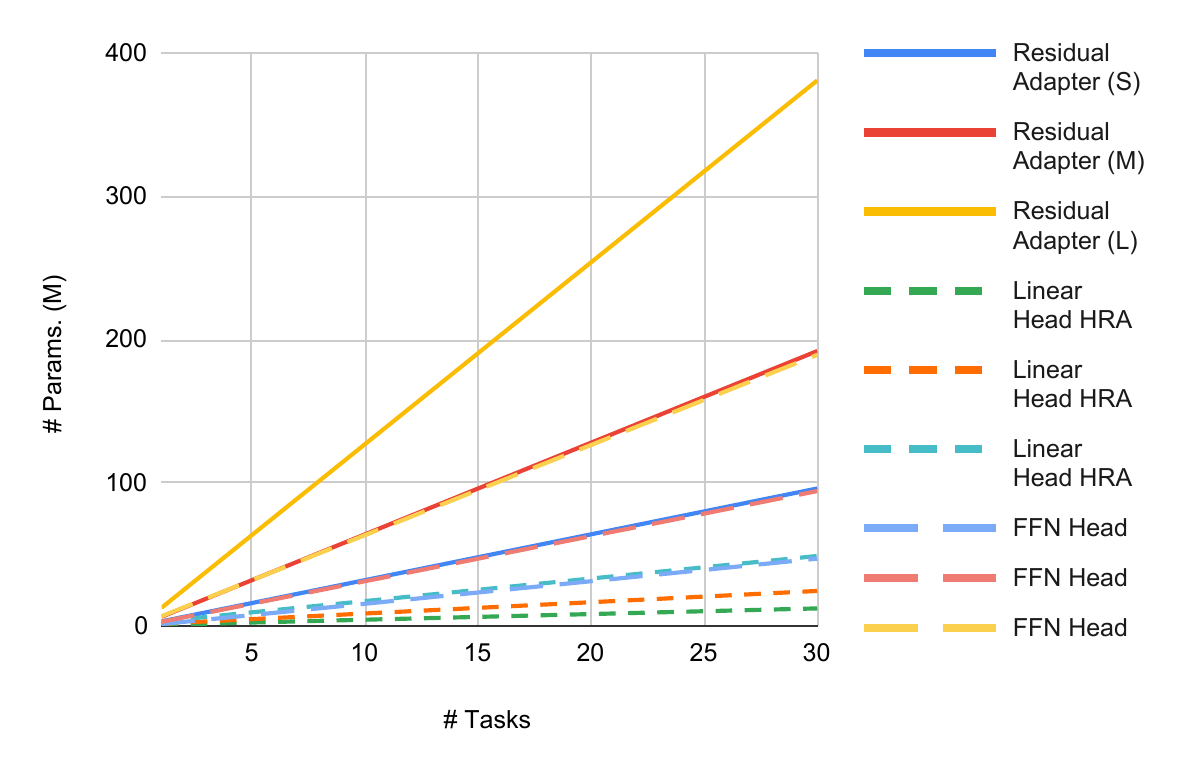}}
    \end{minipage}
    \caption{Our HRA is more parameter efficient with increasing number of tasks while obtaining improved WER performance. Depending on the adapter size, some adapters have sub-linear trend in parameter efficiency.}
    \label{fig:param_task}
\end{figure}

One major advantages of using the one-hot-embedding is that most of the trainable adapter parameters are independent across speakers, resulting in 128 times training throughput efficiency for multi-task adapter experiments. We observe the FFN Head HRA with 201M total parameters achieves the best WER when compared against Residual Adapter, even more close to the golden baseline (full model fine-tuning).

Figure~\ref{fig:param_task} shows the growth rate of the model size when the number of tasks increase. It is observed that FFN Head HRA has a sub-linear growth in terms of the size of adapter parameters with an increasing number of tasks.

\subsection{Online Adaptation}

Table~\ref{tab:multi-task-online-adaptation} reports the WER results from our multi task adaptation experiments with and without pre-trained controller. We hand picked an extra 128
Euphonia
% dysarthric
speaker data as out-of-domain data with respect to the in-domain 128
Euphonia
% dysarthric
speaker data mentioned above. We divide the training into two steps. First step, we pre-train the recurrent controller network with out-of-domain data. Second step, we freeze the recurrent controller network, use in-domain data to train the adapter head with random initialization. So the number of actual training parameter is reduced in this setup as we only train the adapter head. Furthermore, this approach provides a solution for sensitive data sets that cannot be trained within one model. If we pre-train the recurrent controller network only on non-Personal Identifiable Information (PII) data, and parameterize the adapter head by speaker, then no speaker needs to share tuning parameters with others.

\begin{table}[t]
  \centering
    \caption{Online adaptation WER results on
    Euphonia
    % dysarthric speech
    data sets. Our FFN Head HRA (S) with pre-trained controller achieves comparable results against the regular setup (only 0.2\% WER loss). Paired T-Test shows no statistically significant difference between with and without pre-trained controller.}
    \label{tab:multi-task-online-adaptation}
    \begin{tabular}{l c c c}
        \toprule
        \multirow{1}*{Model} & \# Params. & Mean & Paired T-Test \\
        \toprule
        \multirow{3}*{Linear Head HRA}
        & 51M & 10.6 & - \\
        & 102M & 10.9 & - \\
        & 203M & 11.0 & - \\
        \midrule
        \multirow{3}*{FFN Head HRA}
        & 201M & \bf 9.9 & - \\
        & 403M & 10.2 & - \\
        & 806M & 10.4 & - \\
        \midrule
        \midrule
        \multirow{3}*{\shortstack[l]{Linear Head HRA \\ (w/ pre-trained \\ controller)}}
        & 51M & 10.7 & 0.59 \\
        & 101M & 11.0 & 0.25 \\
        & 202M & 11.3 & 0.03 \\
        \midrule
        \multirow{3}*{\shortstack[l]{FFN Head HRA \\ (w/ pre-trained \\ controller)}}
        & 118M & \bf 10.1 & 0.17 \\
        & 269M & 10.3 & 0.14 \\
        & 672M & 10.5 & 0.22 \\
        \bottomrule
    \end{tabular}
\end{table}

\subsection{Model Ablation}
Our Linear Head HRA is structurally similar to Residual Adapters. We can obtain Residual Adapters with shared weights by removing the recurrent states of the RNN controller and then further by unshared the weights, we recover the original Residual adapters. In Table~\ref{tab:abl_res}, we listed the performance for each of the model variants. Removing the recurrent state resulted in a small regression in WER while unshared weights on top of it improved performance but now the number of parameters is more than 100M.

We have also performed an ablation on controller RNN architecture. In addition to the IndRNN, we run benchmarks on the standard RNN with $tanh$ activation and Light GRU~\cite{ravanelli2018light} as controller. The results are summarized in Table~\ref{tab:abl_ctr}. IndRNN and Light GRU both are competitive whereas the RNN with $tanh$ activation underperformed. This confirms that the choice of controller architecture is crucial in our HRA adapters.

\begin{table}[t]
    \centering
    \caption{Linear Head HRA ablation results.}
    \label{tab:abl_res}
    \begin{tabular}{l c c c}
        \toprule
        Model variant & \# Params. & VS & VS w. PN \\
        \toprule
        Linear Head HRA
        & 3.2M &  5.7    &  16.7 \\
        \hspace{0.1cm} - Recurrent state
        & 3.2M &  5.9    &  16.8 \\
        \hspace{0.4cm} - Weight unshared
        & 102.4M &  5.3    &  15.5 \\
        \bottomrule
    \end{tabular}
\end{table}

\begin{table}[t]
    \centering
    \caption{Recurrent controller ablation results.}
    \label{tab:abl_ctr}
    \begin{tabular}{l c c c}
        \toprule
        Controller variant & \# Params. & VS & VS w. PN \\
        \toprule
        IndRNN
        & 1.6M &  6.0    &  16.9 \\
        RNN
        & 1.9M &  6.1    &  17.1 \\
        Light GRU
        & 2.4 &  6.0    &  16.9 \\
        \bottomrule
    \end{tabular}
\end{table}
\section{Conclusion}
\label{sec:conclusion}

We presented Hierarchical Recurrent Adapters (HRA).
By defining a concept of task-level adapter head in HRA, we allocate a shared single adapter controller for all tasks while allowing an individual adapter head to specialize for a new task. This reduces the per-task adapter parameter overhead and enables more efficient adaptation training and inference. The proposed HRA demonstrated better WERs with 2-8x less parameters in single as well as multi-task evaluations. Furthermore, The HRA outperformed the full fine-tuning baseline, at only 12.8M parameters.

% \section{Acknowledgements}
% \label{sec:acks}

% We thank ... for helpful discussions and feedback.

% \vfill\pagebreak

\bibliographystyle{IEEEtran}
\bibliography{refs}

\end{document}